# Universal Arduino-based experimenting system to support teaching of natural sciences

**Z Gingl**[1], **J Mellár**[1], **T Szépe**[1], **G Makan**[1], **R Mingesz**[1], **G Vadai**[1] **and K Kopasz**[2]

*Department of Technical Informatics, University of Szeged, Árpád tér 2, 6720 Szeged, Hungary*

*Department of Optics and Quantum Electronics, University of Szeged, Dóm tér 9, 6720 Szeged, Hungary*

**Abstract.** The rapid evolution of intelligent electronic devices makes information technology, computer science and electronics strongly related to the teaching of natural sciences. Today almost everybody has a smart phone that can convert light, temperature, movement, sound to numbers, therefore all these can be processed, analysed, displayed, stored, shared by software applications. The fundamental question is how education can follow this knowledge and how can education take its advantages. Components and methods of modern technology are available for education also, teachers and students can play with parts and tools which were previously used only by engineers. A good example is the very popular Arduino board which is practically an industrial microcontroller whose pins are wired to easy-to-use connectors on a printed circuit board. In this paper we show a universal system which we have developed for the Arduino platform to support experimenting and understanding of the most fundamental principles of the operation of modern devices. We show our related educational concept and discuss the most important features of the system. Open source hardware and software are available and we provide a number of video tutorials as well.

## 1. Introduction

*1.1. Modern technology and education*

Most of the devices around us are based on electronics and operated by software running on its embedded processor. These systems use sensors to translate real-world signals to numbers that allows information processing and acting on the real-world objects is also possible. Home appliances, medical devices, industrial and commercial robots, smart phones, autonomous cars represent a wide range of examples. It is clear that education must be influenced by this in various ways. However, it seems to be rather difficult to follow the rapid development of technology.

*1.2. Universal principles*

On one hand, there are many modern tools to support education including components to build a robot or some other amazing gadget very easily. On the other hand, it is not evident that students can understand such a modern world around them, one may think that there are even more "black boxes" in education than before. However, the main principles can be surprisingly universal and similar in distant fields. Extracting and applying these can help to develop comprehensive and confident knowledge in an exciting and inspiring practical learning environment.

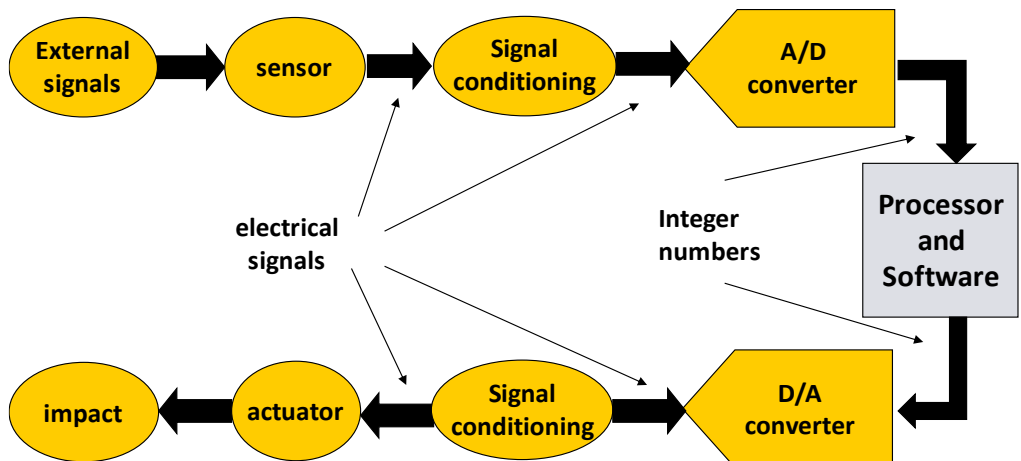

**Figure 1.** Main components of a general modern electronic device

Figure 1 shows the main components of a modern device. For instance, in a smart phone the sound can be sensed by a microphone. Its output signal is handled by some electronics (signal conditioning) to generate a voltage that can be converted to a sequence of integer numbers using an analogue-to-digital converter (ADC). These numbers can be processed by software in various way. One example is the speech recognition common in many smart phone apps. The signal chain can be reversed as shown in the lower part of figure 1. The result of information processing produces numbers that can be converted to an analogue signal using a digital-to-analogue converter (DAC). Its output is fed into a circuitry that can drive the so-called actuators (like a speaker) to provide real-world impact. The device can talk, vibrate or generate light. It is easy to see, that similar principles apply to living things. Humans can see, so sense light and can process the information. As a result of thinking, they can make movements or emit sounds. Software operated devices are often called smart: they are "taught" by programmers. Indeed, it is similar to the education process.

*1.3. Modern tools in education*
There are many tools that are developed to support the application of modern technology (sensors, actuators, digitizing, programming) in education including Lego Mindstorms robots [1], Raspberry PI [2], Arduino [3], BBC micro:bit [4], Scratch by MIT [5]. They can be used at different levels of education and in different disciplines to build intelligent gadgets and to teach the basics of coding and signal processing. However, many things can remain hidden, since the building blocks are not transparent. STEM education may need significantly more to let the students understand the operation and application possibilities much better. From this point of view Arduino is one of the most suitable building blocks due to its simplicity, transparency and relatively low level of integration.

**2. Concept of using modern tools in education**
The Arduino platform is widely used in physics education. It has low cost and many additional components are available. A lot of application examples can be found in the papers, including building devices to measure voltage [6], temperature [7], pressure [8], time interval [9] and many more in various experiments to teach physics attractively.

*2.1. Problems*
Since there are a very wide range of Arduino applications published in the Internet, it is easy to find a solution for almost any task. It is rather simple to reproduce these without considerable efforts, therefore students might not be motivated to create their own version and to understand the basics of operation. Additionally, this can also hide important working principles as the abovementioned integration does.
On the other hand, most of the solutions are not developed by experts, therefore the standards and rules of correct application of the parts are often violated. This is a typical problem in many fields: the large community of the Internet creates a lot of answers that can be very useful, but it can be hard to pick a surely correct one from a huge selection. Therefore, the reliability can be rather low. Questionable solutions can spread very quickly and it often seems hopeless to correct common misunderstandings. In conclusion, development of the right attitude, critical approach, reliability and carefulness may not be supported despite these are essential in high quality STEM education.

*2.2. Engineering-like approach*

Our concept is to make the most important features of the Arduino application possibilities as clear as possible. We try to balance an engineering-like and common approach in order to help students and teachers to freely but wisely use parts and tools previously employed only by engineers. We aim to show the most important standard rules and principles in the simplest and most understandable form [10-14]. The basics of practical electronics, instrumentation, coding and processing are typically covered. We also provide thoroughly developed hardware and software solutions. One of these is introduced in the following.

**3. The EDAQino system**

The Arduino platform is used to implement various instruments as we have mentioned already. Those instruments are very transparent; it is easy to identify the elements shown in figure 1. We have developed an additional circuit called EDAQuino shield (plug-in board) that makes the Arduino board more complete and more universal. Our solution combines almost all of the Arduino-based instruments into a single system that can be used for many different measurements, only the sensors have to be replaced for another application, see figure 2.

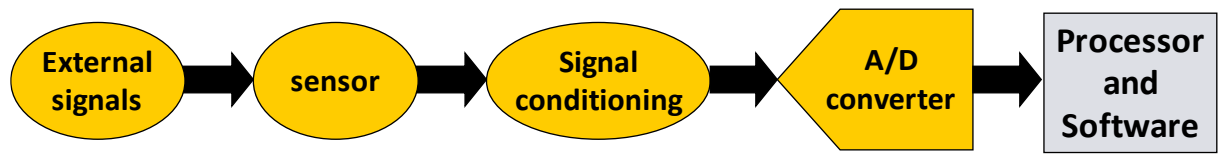

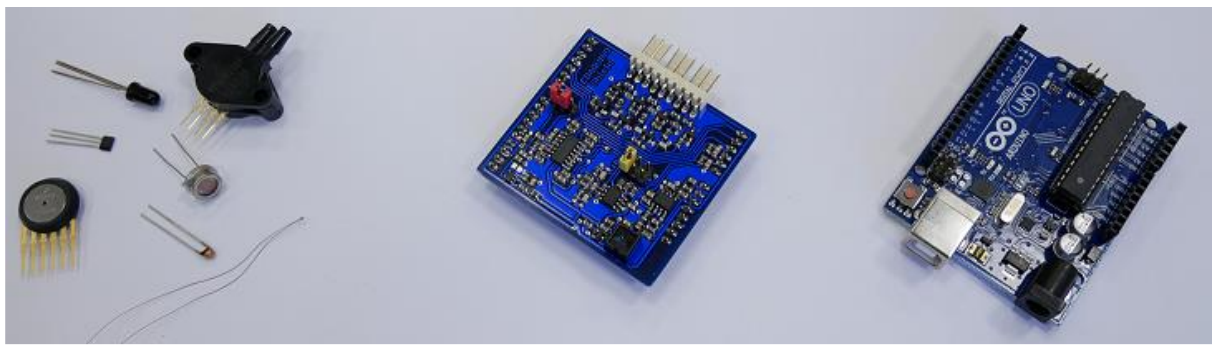

**Figure 2.** Sensors on the left, the universal signal conditioning EDAQuino shield in the middle and the Arduino board on the right form a general instrumentation device.

*3.1. Hardware*

The EDAQuino shield provides similar universal sensor interfacing as its predecessor [15]. Figure 3 shows the simplified structure of the hardware and the electronic equivalent of various sensors. The universal circuitry contains software configurable components to allow direct connection of many different kind of sensors to the three input ports. Voltage, voltage difference, current and resistance can be measured by programming the states of the switches properly. The programmable gain amplifier (PGA) supports the measurement of very low voltage differences; the resolution can be as low as 3.3 uV. Therefore, thermocouples, pick-up coils and Wheatstone bridges can also be connected without any additional circuitry. An infrared photodetector is also integrated and can be used as a photogate or as a sensitive photoplethysmograph to monitor the blood pressure changes in the finger [14]. The documentation is available openly [16] and it can serve as a tutorial on signal condition principles and solutions. If the full universality is not required, some parts of the circuit can be easily assembled on a breadboard or on a pre-drilled copper clad prototyping board. Figure 4 illustrates the implementation of two sensor interfaces. A light-dependent resistor and a thermistor is used in a voltage divider configuration to measure the resistance of the sensors. The user gets information about light intensity and temperature via the resistance measurements.

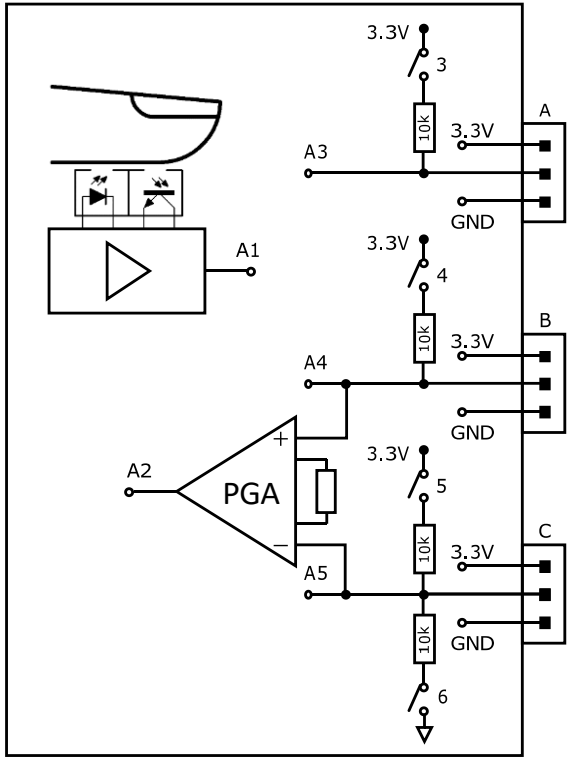

**Figure 3.** The structure of the EDAQuino shield and the electronic equivalents of various sensors. The three input ports can be configured by software. Labels 3 to 6 at the switches are the names of the pins of the Arduino board which are used to drive the switches. A1 to A5 correspond to the analogue inputs.

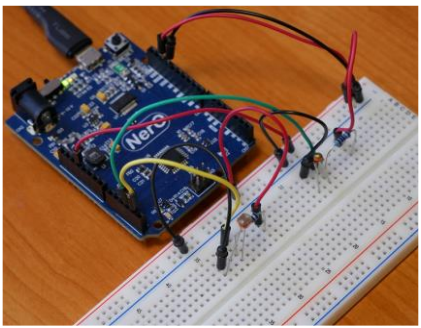

**Figure 4.** Breadboard implementation of the resistance measurement parts of the universal signal conditioning circuitry. A light-dependent resistor and a thermistor is used in a voltage divider configuration to measure the resistance of the sensors. The user gets information about light intensity and temperature via the resistance measurements

*3.2. Software*

We have developed two software components for the measurements. One is running on the Arduino board's microcontroller and the other is executed on the user's computer. The two applications communicate with each other and share the tasks of the measurement. Both are kept as simple and as universal as possible. The user should know how to configure the input ports for a given sensor and how to handle the data. Only the most general functions are supported by the application. Therefore, the students are forced to learn about the most important principles of using sensors, signal conditioning, sampling and signal processing. We intentionally omitted the implementation of automatic detection of sensors and many other functions that could make the use more comfortable.

We have tried to find the optimal balance between the provided functionality and user efforts required for a given measurement task. We have made several video tutorials to help not only the use of the system but to teach the basic principles too [17].

### 3.2.1. Microcontroller code
The microcontroller code accepts simple commands from the host computer. These include configuration of the input ports, setting the sampling rate, starting and terminating the sampling process. Once a measurement has been started, the microcontroller begins to stream the sampled data for all three channels towards the host computer. The code is openly available and due to the permissive license it can be modified to make it compatible with different host applications [16]. It can also be used as a reference, mainly because many questionable solutions have spread for implementing precisely timed sampling [14].

### 3.2.2. Instrumentation application
The application running on the computer is available as an executable and provides the user interface. It is compatible with the original EDAQ530 device [15], only the communication speed had to be changed. The software serves as a real-time chart recorder for the selected input channels and allows to display the data numerically too. Figure 5 and figure 6 show two screenshots to illustrate these features. On the left hand side, a photoplethysmograph signal is displayed which is approximately proportional to the blood pressure changes in the finger [14]. The thresholds are used as event detectors and the time difference between successive heart beats can be displayed in real time. Note that the same functionality can be used generally, for example to measure the period of a pendulum or mass on a spring. On the right hand side of figure 5 and figure 6 the output of a meteorology application can be seen. The temperature, relative humidity and ambient pressure are measured by sensors.

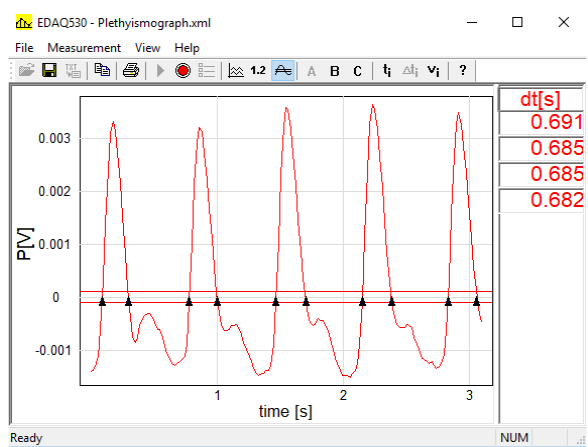
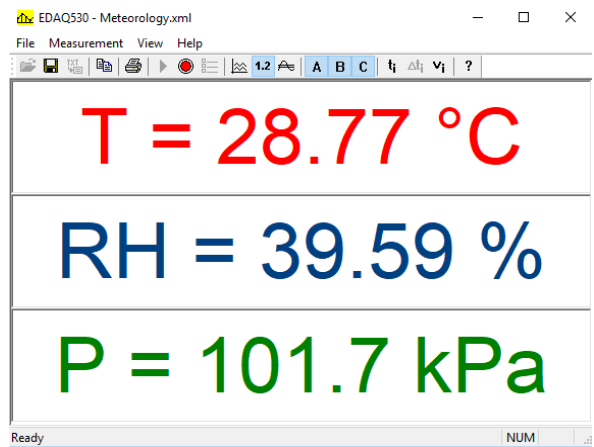

**Figure 5.** Screenshot of the measurement of a photoplethysmograph signal. The level crossing event detection is used to display the time between successive heart beats.

**Figure 6.** Screenshot of the numerical display of the ambient temperature, relative humidity and pressure.

The user has to configure the hardware of all three ports properly and has to provide the functions to scale the measured voltage to the quantities to be displayed. This is shown in figure 7.

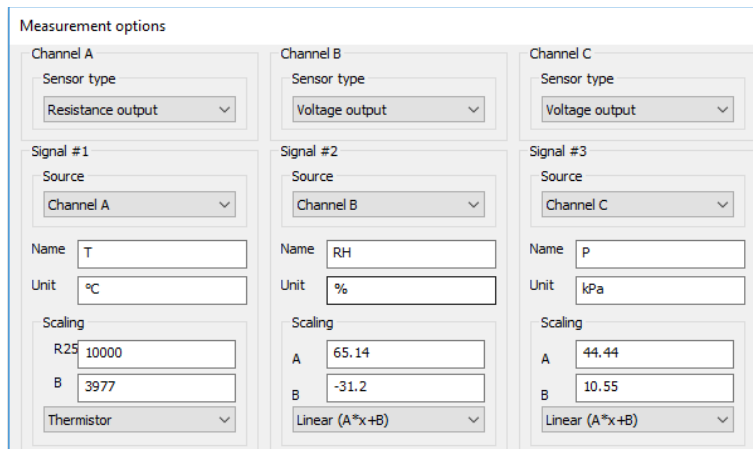

**Figure 7.** Screenshot of the measurement options dialogue box. The user has to select the type of the sensor output for the input ports and the functions to scale the measured voltage to the quantities to be displayed must be given too. Names and units of the quantities are used during numerical display.

## 4. Example experiments
In the following we describe some example experiments that we have tested and made available in video demonstrations as well [17]. These cover some of the options only, there are many ways to work out extensions. The methods shown here can inspire the students to play with the possibilities, to realize their own ideas. Understanding the operation of other education tools may also be helped.

*4.1. Temperature measurement*
The simplest way to measure temperature is to use a thermistor [7]. The user has to configure the port as a voltage divider to measure resistance and has to tell the software that the thermistor equation should be used. Basic knowledge about thermistor characteristics (the resistance at room temperature and the B material constant) is needed. The chart recorder shows spectacularly how the temperature changes in time (see figure 8).

*4.2. Acceleration measurement*
Three axis acceleration sensors (a.k.a. accelerometers) are used in all smart phones to detect tilt by measuring the acceleration due to gravity and to detect movements, free fall as well. Analogue output sensors provide voltage as a linear function of the acceleration, therefore the three signals can be easily measured by the EDAQuino system. The user has to know the offset and gain of the sensor in order to apply the proper scaling function. Accelerometers can be used to investigate tilt, free fall, motion of a pendulum or mass on a spring (see figure 9).

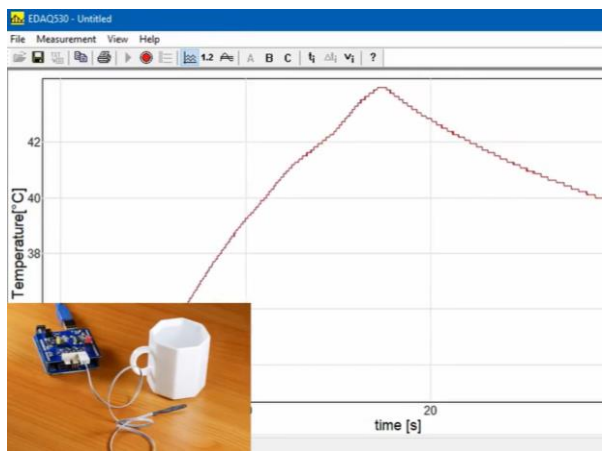
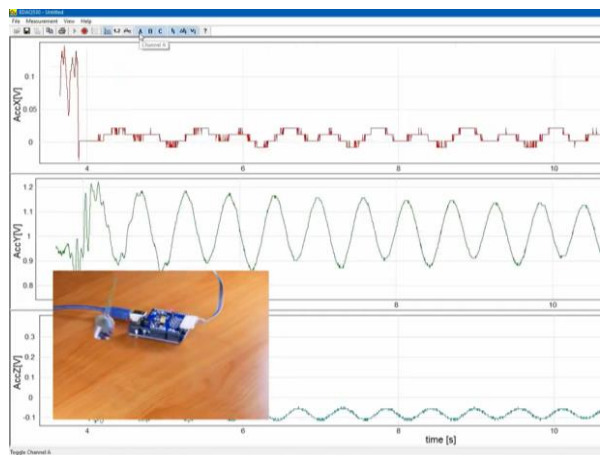

**Figure 8.** Temperature changes can be easily monitored by the use of a thermistor.

**Figure 9.** Signals of a three-axis accelerometer mounted on the bob of a pendulum.

*4.3. Transmissive photogate application*
The photogate is a very useful tool in the teaching of mechanics [18]. In our experiment we have used an infrared LED powered from the port and an infrared phototransistor mounted on a U-shaped plastic bar to implement a transmissive photogate. The output signal of the photogate is low while the object blocks the light between the LED and the sensor, therefore the events associated with the signal changes can be detected. The time instants of the events can be used to determine the time between successive blockings and also the duration of the blocking. Therefore, period and speed can be easily calculated and displayed in real time, what is really attracting (see figure 10).

*4.4. Magnetic field sensor application*
Integrated Hall-sensors output a voltage proportional to the magnetic field at the sensor (see figure 11). These sensors are extensively used to monitor rotation, proximity, periodic motions in various systems. This sensor is powered from the port of the EDAQuino board. It can be used in a similar way as the photogate to detect an object's position as a function of the time. Pendulums, masses on springs and other movements can also be monitored.

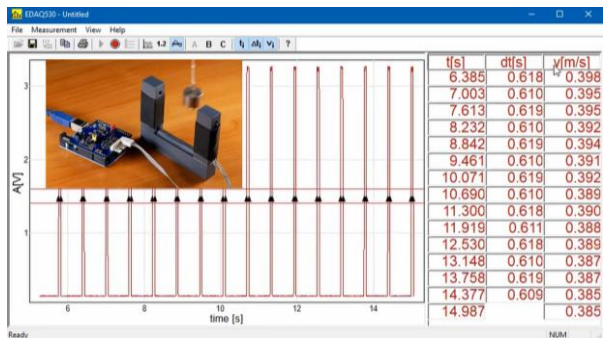 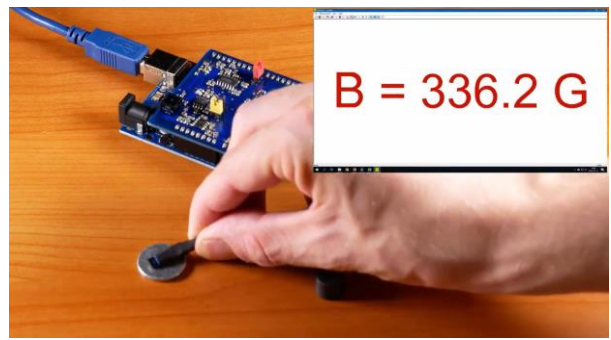

**Figure 10.** Level crossings of a photogate signal can be used to measure the period and speed of a bob on the pendulum.

**Figure 11.** The Hall sensor can be used to measure magnetic field.

*4.5. Heart beats observed by photoplethysmography*
The EDAQuino board has an integrated infrared photodetector. Its sensitivity is very high therefore it can be used to follow small movements or even the blood pressure changes in the finger. The LED's light reflected from the finger depends on the blood density and causes voltage changes at the output of the photodetector. The corresponding signal is shown in figure 5. The built-in level crossing algorithm can be used to detect heart beats and to calculate and display the time difference between successive beats in real-time (see figure 12).

*4.6. Mass on a spring and pendulum*
The infrared photodetector part of the EDAQuino board can also be assembled in a breadboard [14]. This sensitive sensor can be used to follow motion of objects as well. Examples include the monitoring of the oscillations of a mass hanging on a spring or on a cord (see figure 13). The signal shape is rather complex in these cases, but the level crossing detection allow quite accurate measurement of the period.

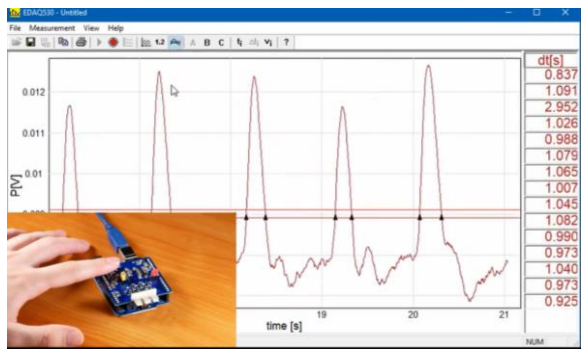
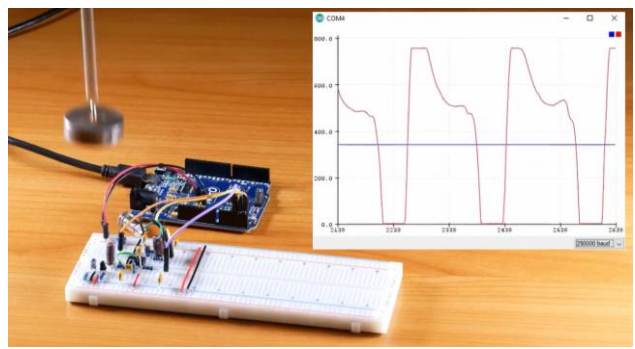

**Figure 12.** Plethysmograph signal and real-time measurement of time between heart beats.

**Figure 13.** Photodetector part of EDAQuino built on a breadboard. It can be used to monitor the motion of a mass on spring.

**5. Conclusion**

In this paper we have shown the results which were presented in a talk at the conference of this proceedings [19]. We have developed an Arduino-based experimenting system that applies and teaches the most fundamental principles of modern devices in a transparent way. The system adds a special sensor interfacing circuitry (EDAQuino shield) plus software to the Arduino board to make it more complete and more universal based on our previous educational measurement system [15]. Our solution combines almost all of the Arduino-based instruments into a single system that can be used for many different measurements, only the sensors have to be changed. We focus on applying an engineering-like educational approach to draw the attention to the most important standard rules and principles in the simplest and most understandable form. Our educational system effectively supports the development of the right attitude, critical approach, reliability and carefulness of the students. These are essential in high quality STEM education and for the engineers and scientists of the future. The hardware documentation and the Arduino source code is openly available and can serve as an educational material. Due to the permissive license these can be modified or some of the parts can be used in own projects. We have made a number of video tutorials to help not only the use of the system but to aid teaching the basic principles as well. Up to now we have given 27 EDAQuino systems to high-schools to support experimentation. Involving the cooperating teachers, we plan to accomplish a study of the attitude of high-school students regarding STEM learning and publish the results in a subsequent paper.

**Acknowledgements**
This study was funded by the Content Pedagogy Research Program of the Hungarian Academy of Sciences.